\input harvmac
\input epsf
%
%
%
%
%
 
\parindent=0pt
\Title{SHEP 96-36}{\vbox{\centerline{Three dimensional massive scalar 
field theory and}\vskip2pt\centerline{the derivative expansion of 
the renormalization group}}}

\centerline{\bf Tim R. Morris}
\vskip .12in plus .02in
\centerline{\it Physics Department}
\centerline{\it University of Southampton}
\centerline{\it Southampton, SO17 1BJ, UK}
\vskip .7in plus .35in

\centerline{\bf Abstract}
\smallskip 
{\sl We show that non-perturbative fixed points of the exact 
renormalization group, their perturbations and corresponding massive field 
theories can all be determined directly in the continuum  -- without
using bare actions or any tuning procedure.} As an example,
we estimate the universal couplings  of the non-perturbative three-dimensional
one-component massive scalar field theory in the Ising model universality 
class, by using a derivative expansion (and no other approximation).
These are compared to the recent results from other methods.
 At order
derivative-squared approximation, the four-point
coupling at zero momentum is better determined by other methods, but
factoring this out appropriately, all our other results are in very close
agreement with the most powerful of these methods.
 In addition we provide for the first time, estimates of 
the $n$-point couplings at zero momentum, with $n=12,14$, and 
the order momentum-squared parts with  $n=2\cdots10$.

\vskip -1.5cm
\Date{\vbox{
{hep-th/9612117}
\vskip2pt{December, 1996.}
}
}
\catcode`@=11 
\def\slash#1{\mathord{\mathpalette\c@ncel#1}}
 \def\c@ncel#1#2{\ooalign{$\hfil#1\mkern1mu/\hfil$\crcr$#1#2$}}
\def\lsim{\mathrel{\mathpalette\@versim<}}
\def\gsim{\mathrel{\mathpalette\@versim>}}
 \def\@versim#1#2{\lower0.2ex\vbox{\baselineskip\z@skip\lineskip\z@skip
       \lineskiplimit\z@\ialign{$\m@th#1\hfil##$\crcr#2\crcr\sim\crcr}}}
\catcode`@=12 
\def\etc{{\it etc.}\ }
\def\ie{{\it i.e.}\ }
\def\eg{{\it e.g.}\ }
\def\cf{{\it c.f.}\ }
\def\viz{{\it viz.}\ }
\def\aka{{\it a.k.a.}\ }
\def\nonp{non-perturbative}
\def\vphi{\varphi}

\def\epsilon{\varepsilon}
\def\halpha{{\hat\alpha}}
\def\alphap{\alpha^{\rm p}}
\def\gammap{\gamma^{\rm p}}
\def\ins#1#2#3{\hskip #1cm \hbox{#3}\hskip #2cm}

\def\x{{\bf x}}

\def\e#1{{\rm e}^{#1}}
\def\E{{\cal E}}
\def\D{{\cal D}}
\def\ss{\mskip 1mu}  
\parindent=15pt

\newsec{Introduction} 

The motivations for the present paper are three-fold. Firstly, we 
wish to expand upon
the renormalization group reasons, given in ref.\ref\hh{T.R. Morris, 
Phys. Rev. Lett. 77 (1996) 1658.},
 for the quantization of renormalised 
couplings. Since these couplings
correspond to relevant (and marginally relevant)
 perturbations about a fixed point, this
necessitates first briefly reviewing how the renormalization group also
`self-determines' the fixed point structure 
itself \ref\deriv{T.R. Morris, Phys. Lett. B329 (1994) 241.}
--\nref\trunc{T.R. Morris, Phys. Lett. B334 (1994) 355.}\nref\revi{T.R. 
Morris, in {\it Lattice '94}, Nucl. Phys. B(Proc. Suppl.)42 
(1995) 811.}\ref\revii{T.R. Morris, 
in {\it RG96}, 
SHEP 96-25, hep-th/9610012.}: 
as we recently emphasised\revii, the fixed-point
and its eigen-operator spectrum -- 
and hence the corresponding massless quantum field theory, can be 
deduced and computed entirely from consistency arguments applied to the 
effective action  and its exact renormalization group 
flow close to the fixed point -- \ie
without needing to go through the construction of introducing 
an overall cutoff $\Lambda_0$, a 
sufficiently general bare action, and then taking the continuum
limit $\Lambda_0\to0$. Thus the universal continuum properties are
accessed directly without this standard, but for quantum
 field theory, actually artificial and extraneous,
scaffolding. 

It would seem rather strange if having been able in this way to derive
directly the continuum massless theory,
the corresponding massive theory required the reintroduction of the
concepts of bare couplings and corresponding tunings to reach the
continuum limit. At first sight this appears to be the case, however
the second and main purpose of the present paper is to
show that once the massless theory's
fixed point and eigen-operator spectrum are known, the massive 
theory may be constructed, again directly, without recourse to any
limiting procedure.  An important byproduct of this analysis, 
is the transformation of the continuum limit of the Wilsonian RG
(Renormalization Group)
into the form of a self-similar flow for the underlying relevant
couplings, in close analogy to
 the usual field theory perturbative
RG, although here the $\beta$ functions are  defined
 non-perturbatively.\foot{Some related remarks can be found in 
refs.\ref\erg{T.R. Morris, Int. J. Mod. Phys. A9 (1994) 2411.}--\nref\wiec{
C. Wieczerkowski, MS-TPI-96-03, hep-th/9603005.}\ref\bon{
M. Bonini {\it et al}, UPRF 96-464, IFUM-525-FT, hep-th/9604114.}. }

We concentrate on a description within the
derivative expansion approximation to the renormalization 
group\deriv\ (see also \ref\twod{T.R. Morris, Phys. Lett. B345 (1995) 
139.}--\nref\ui{T.R. Morris, Phys. Lett. B357 (1995) 225.}\ref\truncm{T.R. 
 Morris, Nucl. Phys. B458[FS] (1996) 477.}), but point out
at the appropriate points how and why precisely the same effects 
can be expected to work in
determining the fixed points, eigen-operator spectrum and massive
continuum limits also for the exact renormalization group. 

Finally, we apply these concepts to a calculation, which however is
interesting in its own right: the universal coupling constant ratios
of the three dimensional Ising universality class scaling equation of state.
In quantum field theory terms, this corresponds to determining the 
 one-particle irreducible $n$-point functions at zero momentum (equivalently
the $O(p^0)$
terms in a series expansion in powers of the momenta)  for 
the three-dimensional one-component massive scalar field theory
based about the \nonp\ Wilson-Fisher fixed point. (These are universal
once they are divided by appropriate powers of the scalar's mass.)
We compute these universal ratios both for the Local Potential Approximation
of the Wegner-Houghton equation\ref\wegho{F.J. Wegner and A. Houghton, 
Phys. Rev. A8 (1973) 401.}\ref\nico{J.F. Nicoll, T.S. Chang and 
H.E. Stanley, Phys. Rev. Lett. 33 (1974) 540.}, and for a smooth
cutoff as utilised in refs.\deriv\twod. The latter allows also to go
to next order -- $O(\partial^2)$ -- in the derivative expansion. We will
see that the results improve, and compare well
with the very recent high order perturbation theory
results of Guida and 
Zinn-Justin\ref\GZJ{R. Guida and J. Zinn-Justin, 
SPhT/96-116, hep-th/9610223.}. Indeed, appropriate universal higher coupling
constant ratios  lie within the errors of
the corresponding resummed perturbation theory results\GZJ. Moreover,
we are able to give estimates for even higher point couplings ($n=12\cdots14$),
and also to estimate the corresponding universal $O(p^2)$ terms in the 
first six $n$-point functions. At present, these
are not available from any other method.

Computing the $n$-point couplings of the scaling equation of state
 in this way is actually a slightly eccentric 
use of the derivative expansion approximation, since the results
most naturally present themselves as full numerical solutions for 
the
equation of state. Indeed in ref.\ref\wet{N. Tetradis and C. Wetterich, 
{Nucl. Phys.} B422 (1994) 541\semi
J. Berges, N. Tetradis and C. Wetterich,
{Phys. Rev. Lett.} 77 (1996) 873.}, a numerical
computation of the scaling equation of 
state\foot{See ref.\ref\Tsyp{
M.M. Tsypin, Phys. Rev. Lett. 73 (1994) 2015.}\ 
for a lattice field theory
computation.}\ has already been presented.
These authors obtain the equation of state
 by tuning an appropriate bare action,  and solving the RG within
 a modified $O(\partial^0)$ approximation 
 incorporating
anomalous scaling (\viz $\eta$).
 Our purpose for computing instead the 
corresponding Taylor expansion coefficients, was primarily
to provide a direct comparison of the first two orders of the
 derivative expansion {\it per se} with
the emerging accurate resummed perturbation theory results, preliminarily
announced by Zinn-Justin, and some independently also by 
Sokolov\ref\sok{A.I. Sokolov, V.A. Ul'kov and E.V. Orlov, to appear
in {\it RG96}, and J. Phys. Studies\semi
A.I. Sokolov, Phys. Solid State 38 (1996) 354.
}\ at the 
August conference ``RG96'' in Dubna. For this comparison, it is sufficient
to compute only in the symmetric phase. 
(The conceptual points of the paper however apply to either phase.) 
A similar computation in the broken phase would be very interesting and
allow estimation also 
of `amplitude ratios'\wet\ by the derivative expansion.
Alternatively these could be computed from the couplings obtained here,
by analytic continuation 
in the mass\GZJ. Extensions to three dimensional
 $O(N)$ symmetric $N$-component scalar 
field theory, and to 
massive two dimensional theories, in particular the infinite 
sequence of multicritical fixed points, described by one-component
 scalar field theory, seems straightforward,
the formalism to $O(\partial^2)$ for these cases
having already been developed 
\revii\twod\ref\turn{M. Turner and T.R. Morris, in preparation.}.
Also, it is certainly possible to estimate even higher point couplings, 
using a more serious numerical attack than we attempt in this paper.
(All the computations for this paper were performed within the Maple
package.)

A more detailed pr\'ecis of the present paper now follows. 
In sec.2, we set up some
of the notation and recall some salient facts: the extraction of the
Legendre effective potential (and action) from the $\Lambda\to0$ limit
of the Wilson effective action -- which is a necessary step to 
construction of the equation of state from the exact RG,
the self-determination of the 
fixed point structure through the requirement of non-singularity,
the large field behaviour and thus determination of the fixed point's
Legendre effective potential. We point out that the leading large field
asymptotics are independent of the effective cutoff $\Lambda$, physically a
reflection of the fact that quantum corrections to the free energy are
negligible for sufficiently large fields.

In sec.3, we study the spectrum of perturbations about the fixed point
({\it a.k.a.} eigen-operators). We show that quantization of the
spectrum of dimensions of these operators follows from the requirement
that the scale dependence can be absorbed in an associated coupling,
\ie a renormalised coupling, and thus  correspond
to the universal self-similar flow
characteristic of the continuum limit.
This is not guaranteed beyond the linearised
level. By examining the flow
exactly, which can be done for large fields, we establish that only
perturbations behaving as a power of $\phi$ for large field $\phi$,
can be associated with a renormalised coupling. As a byproduct we 
derive the form of the Legendre effective potential to first order
in the associated physical coupling. While these results are established
within the Local Potential Approximation (LPA), we explain why (as in sec.1)
the same results should follow also for the exact RG. Finally
in this section, we apply Sturm-Liouville theory to the LPA to prove
that those perturbations that do not behave as power-law for large $\phi$,
collapse under evolution with scale into an infinite sum of perturbations 
that do behave as power-law. Thus all continuum physics is described
in terms of the fixed point, the `power-law perturbations', and their
associated renormalised couplings. The main statements of this section
confirm and generalise the corresponding findings of ref.\hh.

The previous two sections thus describe how to obtain {\it directly} any
massless continuum limit, and its associated operator spectrum,
``directly'' because no limit is actually taken.
In sec.4, we determine how to set up directly the 
corresponding massive continuum limits. We first review the standard lore
on how a \nonp\ massive continuum limit is obtained, defining in the
process the so-called renormalised trajectory. We show that this
trajectory can be defined directly in terms of a boundary condition on 
the flow, involving a coupling $g$, such that finite values of $g$ yield
finite continuum limits, and obtain as a consequence the leading terms
in large $\phi$ dependence of the corresponding massive continuum limit. 
The boundary condition still involves a limiting procedure (its dependence
on $\Lambda$ is prescribed as $\Lambda\to\infty$), but we show that
this limit can
also be removed by reexpressing the flow in terms of the underlying 
renormalised couplings and their associated beta-functions. In addition
we set up a change of variables so that the universal ratios 
 of couplings (\viz
physical scale
independent ratios) are obtained once all modes have
been integrated out (\ie as $\Lambda\to0$).
Again in this section, we point out how and why these observations should
hold for the exact RG.

In sec.5, we formulate the theory of sec.4 specifically in terms of the
LPA of the sharp cutoff (\aka Wegner-Houghton) flow equation,
applied to three dimensional one-component scalar field theory
governed by the \nonp\ Wilson-Fisher fixed point, particularly the
universal coupling constant ratios of the Legendre effective potential 
(equivalently scaling equation of state).
We discuss
the numerical implementation of the boundary conditions and flows, expansion
directly in terms of the (ratios of) couplings, and the corresponding choice of
(and independence of) closure ansatz. The short sec.6 delineates the 
differences that arise in using the $O(\partial^0)$ approximation, 
equivalently LPA, of the smooth cutoff flow equation derived in ref.\deriv.

In sec.7 we go beyond the LPA, to order derivative-squared,
thereby also allowing estimation of universal ratios of
the momentum-squared terms in $n$-point Green functions, and more
accurate estimation of the scaling
 equation of state. Wave-function renormalisation
and running anomalous dimensions are dealt with explicitly here. 
Once again we provide here also the explicit formulae needed for the
numerics.

Finally, in sec.8 we present the numerical results. We present numerical
evidence showing independence of closure ansatz, for three different
choices of ansatz. Picking the most rapidly converging ansatz,
we deduce the sharp cutoff LPA,
the $O(\partial^0)$, and $O(\partial^2)$ results for the universal
coupling constant ratios, and combine the
derivative expansion results into one final number together
with an estimated error
of truncation of the derivative expansion. These are compared to
results from resummed perturbation theory\GZJ. The perturbative methods
are more powerful for low order couplings but the derivative expansion
eventually wins out for higher order couplings. Following ref.\GZJ, we 
then factor out an overall normalization (effectively
the size of the four-point
coupling) and present derivative expansion results (and sharp cutoff
LPA results) for the corresponding ratios $F_{2l-1}$. These are seen
to be much more accurately determined by the derivative expansion,
and are in close agreement with the most accurate determinations
from other methods. (A table of comparisons is presented, including the
results from
resummed perturbation
theory, $\epsilon$ expansion, the exact RG approximation of ref.\wet, 
high temperature series and Monte-Carlo estimates.)
We thus conclude that the main error in the derivative expansion is
in determining the lowest point coupling(s), and can be absorbed in
an effective normalisation factor $\zeta_{eff}$.  

There is no need for a separate summary and conclusions, since this has
already in effect been incorporated in the above introduction and pr\'ecis.

\newsec{Fixed Points} 

First, let us review how the fixed points are determined by consistency
arguments alone. To do this, we find it easier to concentrate on a 
specific equation and then to point out how the arguments generalise.
Thus consider the so-called LPA (Local Potential Approximation) \nico\ to the
Wegner-Houghton renormalization group\wegho. We remind the reader that
the latter is just\deriv\erg\ the sharp cutoff\foot{We look first at this
long established version of the LPA. Later we will employ smooth cutoff
versions, which also turn out to be more accurate. The subtleties involved
in handling a sharp cutoff, both technical and physical, have been 
extensively addressed in ref.\truncm.}\
 limit of Polchinski's 
form\ref\pol{J. Polchinski, Nucl. Phys. B231 (1984) 269.}\ of
Wilson's 
continuous RG\ref\kogwil{K. Wilson and J. Kogut, Phys. Rep. 12C (1974) 75.} (Renormalization Group):
$$
{\partial S_\Lambda \over\partial\Lambda}={1\over2}\,{\rm tr}\, 
{\partial\Delta_{UV}\over\partial\Lambda}
\left\{ {\delta S_\Lambda\over\delta\phi}
{\delta S_\Lambda\over\delta\phi}-{\delta^2S_\Lambda\over\delta\phi\delta\phi}
-2\left(\Delta^{-1}_{UV}\phi\right)
{\delta S_\Lambda\over\delta\phi}\right\}\quad,
$$
(where $\Delta_{UV}(q,\Lambda)=C_{UV}/q^2$ is a cutoff massless propagator.
 For more details on this equation see refs.\revii\erg\pol\kogwil.
It will not however be used in the rest of the paper.)
The approximation corresponds to projecting on an effective action
of the form
\eqn\appr{S_\Lambda=\int\!\!d^D\!x\,\left\{\half(\partial_\mu\phi_{\rm phys})^2
+V_{\rm phys}(\phi_{\rm phys},\Lambda)\right\}\quad,}
\ie to discarding all momentum dependent terms generated by the RG.
($D$ is the space-time dimension.)
For a single component scalar field, the result 
is\nico\ref\parelpa{A. Parola and L. Reatto, Phys. Rev. A31 
(1985) 3309.}\ref\hashas{A. Hasenfratz and  P. Hasenfratz, Nucl. Phys. 
B270 (1986) 687.}\trunc\truncm:
\eqn\shar{{\partial\over\partial t}V(\phi,t)+d\phi
V'(\phi,t)-D V(\phi,t)= \ln\left[1+V''(\phi,t)\right]\ \ .}
Here $t=\ln(\mu_p/\Lambda)$, $\mu_p$ is an arbitrary physical 
mass scale, 
the primes stand for differentials with respect to $\phi$, and
$d$ is the scaling dimension of the field $\phi$.
Generally this is given at a fixed point by $d=\half(D-2+\eta)$, however a 
consequence of the LPA is that the anomalous dimension $\eta=0$.
In this equation the scaling dimensions of the field and potential
have been used to construct dimensionless equivalents \trunc:
\eqn\sca{\phi=\phi_p/\Lambda^d\ins11{and}
 V(\phi,t)= V_p(\phi_p,\Lambda)/\Lambda^D \quad.}
Thus $p$-subscripted 
quantities refer to dimensionful $p$hysical 
quantities. This step is equivalent to (but more direct than) 
the traditional rescaling of the cutoff
back to its original size after a blocking\kogwil\wegho.

Actually, a further purely numerical transformation is necessary to
obtain the properly normalised physical quantities appearing in \appr:
\eqn\pret{\phi_p=\zeta\phi_{\rm phys} 
\ins11{and} V_p=\zeta^2V_{\rm phys}\quad,}
where $\zeta=(4\pi)^{D/4}\sqrt{\Gamma(D/2)}$ was chosen to 
prettify the eqn.\shar. This step will only need to be taken into
account in the final sections containing the numerics. 

It is worth recalling here that the effective potential of the Wilsonian
effective action $S_\Lambda$ tends, in the limit $\Lambda\to0$, to
the Legendre effective potential\erg\foot{providing that 
${\partial\over\partial q^2}C_{UV}|_{q=0}=0$, as follows trivially from
observations in 
refs.\erg\truncm\ref\largen{M. D'Attanasio and T.R. Morris, Southampton
preprint SHEP 97-03.}}\
 (\ie generator of one-particle irreducible
Green functions evaluated at zero momentum). The latter is of course
precisely what we will need later, to construct the equation of state.
This is one consequence of the deeper connection also proved in ref.\erg, 
namely that the Legendre effective action (minus its infrared regulated
kinetic term\erg):
$\Gamma_\Lambda$, for a field theory with
infrared cutoff $C_{IR}(q,\Lambda)$ imposed, is a generalised
Legendre transform of Polchinski's Wilsonian effective action $S_\Lambda$
computed with ultra-violet cutoff $C_{UV}(q,\Lambda)=1-C_{IR}(q,\Lambda)$:
\foot{It does not matter here whether quantities are physical,
or dimensionless combinations.}
\eqn\Leg{S_\Lambda[\phi]=\Gamma_\Lambda[\phi^c]+\half 
(\phi^c-\phi).{q^2\over C_{IR}}.(\phi^c-\phi)\quad.}
 (Somewhat similar remarks have been 
made by other authors\ref\apair{M. Salmhofer, Nucl. Phys.
B (Proc. Suppl.) 30 (1993) 81\semi
M. Bonini, M. D'Attanasio and G. Marchesini, Nucl. Phys. B418 
(1994) 81.}\ref\wet{C. Wetterich, Phys. Lett. B301 (1993) 90.}.
{\it N.B.} $\phi^c$ in this equation is the ``classical field'',
 defined as a function of $\phi$ through the Legendre transform, 
in the usual way, \ie by differentiating
\Leg\ by $\phi$ at fixed $\phi^c$.)
In this way we can have our cake and eat it: fixed points in the Wilsonian
RG have a natural physical explanation through thinking about blocking, but
at the same time we can recover the information apparently blocked away --
via this generalised Legendre transform relation.

At a fixed point, $V(\phi,t)= V(\phi)$ is fixed \ie independent of $t$. From
eqn.\shar, it would appear at first glance that there is a two parameter
set of fixed point solutions $V(\phi)$ (because \shar\ reduces to a second
order ordinary differential equation), but this is true only locally.
In fact all but typically a discrete number\foot{an exception being
critical sine-Gordon models in two dimensions\twod}\ of these solutions
are singular at some finite value of the field $\phi$ \deriv\twod\hh\revii. 
Clearly the presence of such a singularity
is unacceptable \eg in leading to violations of Griffiths' analyticity.
From our experience, the solutions that are well-defined for all
real $\phi$ (typically then countable in number), may all be 
identified with approximations to the expected fixed points
of the exact RG, in the sense that they have the
right qualitative properties and yield reasonable quantitative 
answers  \deriv\twod\hh\revii. Importantly there appear to be none of
the `spurious fixed points' that afflict higher order truncations in 
powers of the field\trunc\revii.

The form of the fixed point solution for large $\phi$ may be
ascertained from eqn.\shar\ \trunc:
\eqn\larph{V(\phi)\sim A\phi^{1+\delta}\ins11{as}\phi\to\infty\quad,}
with $A\ge0$ an as yet unkown coefficient.
We have used a hyperscaling relation 
to write $D/d-1$ as the
critical exponent $\delta$. (We are assuming $d>0$ here and from now on.
This thus excludes the two-dimensional exceptions\twod.
All scaling relations 
trivially hold automatically in derivative
expansion approximations since scale invariance is not
broken and the massless limit may be directly considered, as we are 
indeed so doing here.)
The leading term above
remains valid beyond the LPA and for any cutoff
{\sl in the form written}\foot{\ie with $\delta=D/d-1$,
and $d$ the full scaling dimension
following from a non-zero $\eta$.}\
 because it simply
arises from solving the left hand side of 
eq.\shar,\foot{the neglect of the right hand side being 
justified by inspection}\ which in turn is
purely generated by the assignment of scaling dimensions $d$ and $D$ to
the field and potential respectively.
This power law is thus precisely as required so that $V_p(\phi_p,\Lambda)$
has a finite non-vanishing
limit as $\Lambda\to0$. Indeed, from \sca\ and \larph,
we obtain
\eqn\frem{V_p(\phi_p,0)=A\phi_p^{1+\delta}\quad.}
In view of the relation to the Legendre effective potential outlined above, 
this is nothing but the free energy per unit volume, at the critical point,
as a function of the field \trunc.
Notice that the behaviour of $V_p(\phi_p,\Lambda)$ in the
regime $\phi_p>\!>\Lambda^d$, is also given by \frem, and in particular is
thus actually independent of $\Lambda$. Physically, this is precisely to
be expected: it reflects the fact that quantum corrections to the free
energy are negligible for sufficiently large fields. Indeed
the asymptotic behaviour \larph\ for a fixed point, is the only such
behaviour consistent with this `mean-field-like' evolution \ie one in which
quantum corrections are absent.

It seems reasonable to suppose that the same characteristics, 
as outlined above,
also completely determine the fixed points in the exact renormalisation
group. In this case\erg, the fixed point equation is a second order
non-linear functional differential equation and presumably, locally
has a full functional space worth of solutions but globally only 
typically a discrete number which are
well defined for all $\phi(\x)$, these latter being again the only
{\it bona fide} fixed points.

\newsec{The eigen-operators} 

To obtain the spectrum of (integrated scalar) eigen-operators we linearize
around the fixed point by writing $S_\Lambda[\phi]=S[\phi]+
\epsilon\, \e{\lambda t} s[\phi]$, or in terms of the LPA:
\eqn\pert{V(\phi,t)=V(\phi)+\epsilon\, \e{\lambda t} v(\phi)\quad,}
to first order in $\epsilon$. 
Here we have used the fact that the RG is quasilinear in $t$, and used
separation of variables. Thus we have from \shar:
\eqn\sharper{
\lambda v(\phi) + d\phi v'(\phi) -D v(\phi)={v''(\phi)\over 1+V''(\phi)}
\quad.}
This time the ordinary second order differential equation is linear
with non-singular coefficients and therefore we are guaranteed to
have for {\sl every real value of} $\lambda$, 
 a continuous 
one-parameter (up to the arbitrary normalisation)
set of {\sl globally well-defined} real solutions. How can we square this
with the fact that experiment, simulation \etc, 
typically only uncover a discrete spectrum of such operators
and in particular 
only a discrete number of relevant directions corresponding 
to eigenvalues $\lambda>0$? The answer was given in ref.\hh\revii (if rather
compactly): {\it only the discrete set of solutions for $\lambda$ and
 $v(\phi)$, where $v(\phi)$ behaves as a power of $\phi$ for
large field, can be associated with a corresponding renormalised
coupling $g(t)$ and thus universal self-similar flow close to the 
fixed point}. 

This comes about as follows. 
First note that if $v(\phi)$ is to behave as power-law for large $\phi$,
then from \sharper, this power is determined as
\eqn\larpher{
v(\phi)\sim a \phi^{1+\delta-\lambda/d}\ins11{as}\phi\to\infty\quad,}
with $a$ an as yet unkown coefficient.
Again, as with \larph, this remains valid beyond the LPA and for any
cutoff since the power is determined only by the left hand side of \sharper.
Once this power law is imposed for $\phi\to\infty$, and for $\phi\to-\infty$
with possibly different coefficient,\foot{or a boundary condition
following from symmetry considerations, is imposed at the origin\twod\revii}\
and linearity of \sharper\ taken into account, we see that we have
sufficient boundary conditions to overconstrain \sharper\ leading
to typically a discrete set of solutions for the eigen-values
$\lambda$ and eigen-operators $v(\phi)$ \deriv. Therefore we need now to explain
only why we require $v(\phi)$ to behave as a power-law for large $\phi$. 

Studying eqn.\sharper, and using \larph, 
we see that if $v(\phi)$ does not behave as 
\larpher\ then for $A>0$, it must behave as
\eqn\larbad{v(\phi)\sim \exp A(D-d)\ss\phi^{1+\delta}
\ins11{as}\phi\to\infty \quad,}
where we are neglecting a multiplicative power-law correction. Only for
the Gaussian fixed point $V(\phi)\equiv0$, does
$A=0$, and in this case we have instead 
$v(\phi)\sim \exp d\ss\phi^2/2$. The behaviour \larbad\
is changed by going beyond
the LPA, and depends on the choice of cutoff\foot{\eg compare the
formulae of ref.\deriv.}\ because it results from balancing the
right hand side of \sharper\ against the derivative term of the
left hand side. The crucial point for us however is that if $v(\phi)$
does not behave as \larpher, then it diverges faster than power-law
as $\phi\to\infty$. And this statement is presumably universally true,
holding for any (valid) choice of cutoff and also for the exact RG. 

From eqn.\pert, it is tempting to identify $g(t)=\epsilon\e{\lambda t}$ 
as a (renormalised) coupling conjugate to the operator $v(\phi)$,
but this $t$ evolution followed from linearising in $\epsilon$, and
for any finite coupling, no matter how small, there is a regime of
large fields where such linearisation is not justified. [Thus \eg
$\epsilon v(\phi)>\!>1$ for $\phi>\!>(-\ln\epsilon)^{d/D}$ in case \larbad.]
In this regime however, it is straightforward {\sl to solve exactly} for the
$t$ evolution because the right hand side of \shar\ contributes
negligably compared to the at-least-power-law behaviour of the left
hand side, while the left hand side --being linear-- is easily solved.
Thus starting at $t=0$ with $V(\phi,0)=V(\phi)+\epsilon v(\phi)$, we have
\eqn\mf{V(\phi,t)\sim {\rm e}^{Dt}\, V(\phi\,
{\rm e}^{-dt},0)\quad =V(\phi)+\epsilon{\rm e}^{Dt}\,v(\phi\,{\rm e}^{-dt}),
\ins{.5}{.5}{as}\phi\to\infty\quad,}
where we have used \larph\ (and are assuming, here and later,
 that subleading corrections to each term can be neglected).
Indeed, once again, the neglect of the right hand side of \shar\ is just the
statement that quantum corrections are negligible in the regime of
large fields; combining \mf\ and \sca,
\eqn\mp{V_p(\phi_p,\Lambda)\sim \mu_p^D V(\phi_p/\mu_p^d,0)
\ins11{for}\phi_p>\!>\Lambda^d\quad.}
We expect then that \mf\ and \mp\ hold also for the exact RG.
Now, for the power-law perturbations \larpher,
\mf\ simplifies to
\eqn\larpow{V(\phi,t)\sim  A\phi^{1+\delta} +\epsilon{\rm e}^{\lambda t}
a \phi^{1+\delta-\lambda/d}
\ins11{as}
\phi\to\infty\quad,}
which is just the large $\phi$ behaviour of \pert\ --
confirming the identification $g(t)=\epsilon\e{\lambda t}$.
\foot{We checked that at subleading
order in the asymptotic expansion, 
the $t$ dependence can still be entirely absorbed
in $g(t)$.}\
Correspondingly from \mp:
\eqn\plarpow{
V_p(\phi_p,0)= A\phi_p^{1+\delta}+\epsilon\mu_p^\lambda\,
a\,\phi_p^{1+\delta-\lambda/d}\quad,}
from which we see that $g(t)$ corresponds 
to a dimensionful $\mu_p$-dependent physical
coupling $g_p=\epsilon\mu_p^\lambda$ of dimension $\lambda$. But for
the non-power-law perturbations \larbad, and in general for perturbations
rising faster than power-law, the $t$ dependence in \mf\
cannot be combined with $\epsilon$ and thus absorbed into a running coupling.
(Nor can all the $\mu_p$ dependence be absorbed in a physical coupling.)
We have thus confirmed the italicized statement at the beginning of 
this section, and expect that the statement holds beyond the LPA and indeed
also for the exact RG.

Finally, we know from Sturm-Liouville 
theory\ref\Ince{\eg E.L. Ince, ``Ordinary differential
equations'', (1956).}\ applied to 
\sharper\ that {\it as soon as $t>0$}, 
{\it non-power-law perturbations} [\ie behaving as \larbad,  \mf],
{\it can be reexpanded in terms of a sum over the power-law
perturbations. It follows then that in any
case only these are needed to span all the continuum physics.}

To prove these statements, we note that the differential operator $\D$, defined
by writing \sharper\ as $\D\ss v(\phi)=\lambda\, v(\phi)$, is Hermitian
with respect to the weight function 
\eqn\rh{\rho(\phi)=\left\{1+V''(\phi)\right\}
\exp-d\int^\phi\!\!\!d{\tilde\phi}\,
{\tilde\phi}\left\{1+V''({\tilde\phi}\ss)\right\}\quad,}
and the power-law boundary conditions  \larpher\ (imposed at $\phi=\pm\infty$). 
This follows because the large $\phi$ behaviour of $\rho$ 
 is sufficient to ensure that the  bilinear concomitant
(the boundary term generated by integration
by parts) vanishes. Indeed from \larph,
\eqn\lrh{\rho(\phi)\sim \exp-A(D-d)\ss\phi^{1+\delta}
\ins11{as}\phi\to\infty\quad,}
for $A>0$, otherwise $A=0$ and $\rho\sim\exp-d\ss\phi^2/2$\ 
(neglecting again, here and in the ensuing,
 the unimportant multiplicative power-law corrections).
Now, from the general Sturm-Liouville theory,
we have that the power law eigen-functions $v_i(\phi)$ form a discrete set,
$i=1,2,\cdots$, with corresponding, possibly finitely
degenerate, eigen-values $\lambda_i$, such that there is a most
positive eigenvalue, which we call $\lambda_1$, 
and an infinite tower of negative eigenvalues
(\ie taking the $\lambda_i$'s to be non-increasing in $i$, we have
 $\lambda_i\to-\infty$ as $i\to\infty$. Note that these facts concur with
physics expectations). The eigen-functions may be taken
to be ortho-normalised:
$\int^\infty_{-\infty}\!d\phi\ss\rho(\phi)\, v_i(\phi)
\ss v_j(\phi)=\delta_{ij}$.
From \larbad\ and below it, and \mf, we have that a non-power-law
perturbation $v(\phi,t)\equiv V(\phi,t)-V(\phi)$ behaves as
$$\eqalign{A>0: \hskip1cm v(\phi,t) 
&\sim \exp A(D-d)\,\e{-Dt}\ss\phi^{1+\delta}\cr
A=0: \hskip1cm \phantom{v(\phi,t)}&\sim \exp d\,\e{-2dt}\ss\phi^2/2\cr}\ins11{as}\phi\to\infty\quad,$$
thus the expansion
coefficients $c_i(t)=\int^\infty_{-\infty}\!d\phi\ss\rho(\phi)\,
v(\phi,t)\ss v_i(\phi)$, are well defined for all $t>0$, while for
$t> {\ln2\over D}$ ($A>0$, or $t>{\ln 2\over 2d}$ with $A=0$), we have
the completeness relation,
$$\int^\infty_{-\infty}\!\!\!\!d\phi\,\rho(\phi)
\left\{ v(\phi,t)-\sum^N_{i=1} c_i(t)v_i(\phi)\right\}^2
\to0\ins11{as} N\to\infty\quad.$$
We have thus proved the italicized statements of the previous paragraph.
Note that these statements are the generalisation to any fixed point, of
the corresponding statements made about Gaussian fixed points in ref.\hh.

\newsec{Massive continuum limits}

From this detailed study we actually
have in place everything we need to set up {\sl directly} the massive
continuum limit, \ie without actually having to 
perform the limit. Before doing this
however, let us recall the standard lore\kogwil\ on how a \nonp\ 
massive continuum limit is obtained. This is illustrated in fig.1. 
In the infinite dimensional space of bare actions, there is the
so-called 
critical manifold, which consists of all bare actions yielding a given
massless continuum limit. Any point on this manifold -- \ie any such
bare action -- flows under a given RG towards its fixed point; local
to the fixed point, the critical manifold is spanned by the infinite set
of irrelevant operators. Let us assume that emanating out of the fixed
point there is just one relevant direction;  the generalisation to
more than one is straightforward and will be indicated later. 
Choosing an appropriate parametrization of the bare action, we move a little
bit away from the critical manifold. The trajectory of the RG will 
to begin with, move towards the fixed point, but then shoot away 
in the relevant direction (exponentially fast in $t$) towards the so-called
high temperature fixed point which represents an infinitely massive 
quantum field theory. 
\midinsert
\centerline{
\epsfxsize=0.8\hsize\epsfbox{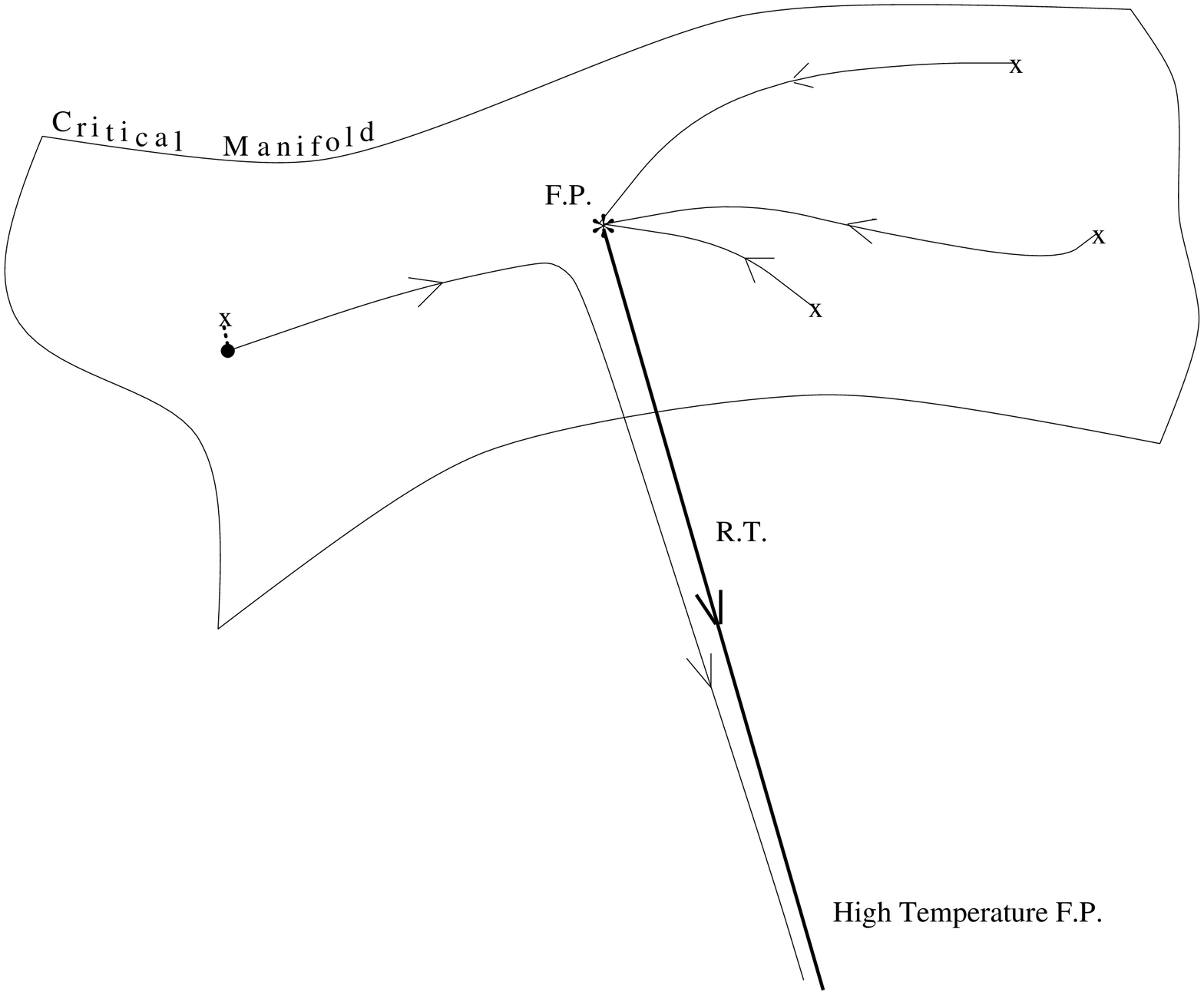}}  
\bigskip
\centerline{\vbox{\noindent {\bf Fig.1.} 
Tuning to a massive continuum limit. x marks points on the critical manifold,
whereas the black blob is a point shifted slightly off the critical manifold.
}}
\endinsert

To obtain the continuum limit, and thus finite
masses, one must now tune the bare action back towards the critical
manifold and at the same time, reexpress
 late $t$ physical quantities in renormalised terms
appropriate for the diverging correlation length. In the limit that the
bare action touches the critical manifold, the RG trajectory splits into
two: a part that goes right into
the fixed point, and a second part that emanates from the fixed point
out along the relevant direction. This path
is known as the Renormalised Trajectory\kogwil\ (RT); the effective
actions on this path are `perfect 
actions'\ref\perf{P. Hasenfratz and F. Niedermayer, Nucl. Phys. 
B414 (1994) 785.}.
In terms of renormalised
quantities, the large $t$ section of this path obtains a finite limit,
namely the effective action of the continuum quantum field theory.

Therefore, to obtain the limits directly we must first describe the RT.
This is given, {\sl close} to the fixed point, by 
\eqn\close{V(\phi,t)=V(\phi)+ g(t) v(\phi)
\quad,}
where the relevant coupling $g(t)$, corresponding to the one relevant direction,
satisfies
\eqn\inig{g(t)=g\e{\lambda t}\quad,}
with $g$ a constant.
Eqn.\close\ is the unique solution that `tracks back' into the fixed point
as $t\to-\infty$. It satisfies the flow equation there [by linearisation 
as in \pert] since $g(t)\to0$. Using \close\ as a 
$t=-\infty$ `boundary
condition' for the flow equation, we may now describe the whole RT.

In a case where there is more than one relevant direction, we 
then of course have a continuum
of possible RT's, which may be labelled by the ratios of the relevant couplings,
thus labelling different physics for given overall mass-scale ({\it a.k.a.}
 correlation length). Eqn.\close\ is replaced by a sum over the relevant
couplings, each of which satisfies an equation of the form \inig\
 as $t\to-\infty$.
These generalizations are straightforward and thus
will not be considered further.

Note now, that any finite value of 
$g$ yields a finite limit for the continuum quantum field
theory. To see this, note that 
since $\lambda>0$, comparison of \larpher\ with \larph\
shows that for {\sl any} finite $t$, there is a 
sufficiently large $\phi$ where the linearised solution \close\ is still
valid! Indeed, the linearisation \sharper\ is valid for \close, 
providing we have
$g(t) v''(\phi)/V''(\phi) <\!<1$, and this
is true for all $\phi>\!> C g^{\beta}\e{d t}$. 
(Here $C=
{a(1+\delta-\lambda/d)(\delta-\lambda/d)\over A\delta(1+\delta)}$,
and we have used the hyper-scaling relation $\beta=d\nu=d/\lambda$.) 
Taking the limits $t\to\infty$ and $\phi\to\infty$, while obeying the
above inequality, we may continue to use \close\ and \inig\ and thus,
similarly to \larpow\ and \plarpow, we obtain the scaling free energy per 
unit volume for large constant fields, for the system
 perturbed from the fixed point:
\eqn\frema{V_p(\phi_p,0)\sim\phi_p^{1+\delta}\left\{A
+a g_p \phi_p^{-1/\beta}\right\}
\ins{.5}{.5}{for}\phi_p>\!> 
C g_p^{\beta}\quad.}
Here we have written $g_p=g\mu_p^\lambda$. The limit is finite
 for any finite $g$, as claimed, at
least in this large $\phi_p$ region, and we see that it corresponds
to a finite physical coupling $g_p$. Furthermore, if $V_p(\phi_p,0)$ 
is finite for all $\phi_p$, we have using \sca, that
 \eqn\smalL{V(\phi,t)\sim V_p(\Lambda^d\phi,0)/\Lambda^D}
should satisfy \shar\ in the limit $\Lambda\to0$. Substituting, we see
that \shar\ is indeed satisfied, in fact
 up to corrections $\sim\ln(\Lambda)/\Lambda^D$.   

We remark that higher order corrections to
\frema\ are presumably directly calculable by continuing the perturbative
expansion \close\ to higher order in $g$, this being determined 
uniquely by similar considerations to sect.3.

Obviously the form of the boundary condition: \close, with \inig\ and $t\to
-\infty$, is rather inconvenient, indeed is yet another limit, and the
flow equation is numerically unstable (stiff) in such a region
-- since perturbations behave exponentially in $t$. We can
resolve both these problems by expressing the 
continuum RG in terms of self-similar flow of the 
renormalised couplings (\ie analogously to the field theory perturbative RG).
In other words we rewrite in this case, \shar\
as a differential equation for $V(\phi,g)$ together with a 
(\nonp) beta function $\beta(g)$ for
$g(t)$.   
The advantage of this is that the flow equation
in $g$ has perturbations
that behave only as a power-law (around the fixed point)
and, as we will see, has remarkably good numerical behaviour. At
the same time we have resolved the limit since the boundary condition 
defining the RT, may now simply be stated as\foot{Similar remarks
have been made in ref.\wiec.}\
$${\partial\over \partial g} V(\phi,g)\Big|_{g=0}=v(\phi)\quad.$$
Rather than furnish further details at this stage, we combine this 
idea with another change of variables.
Since there is really 
only one dimensionful parameter $g_p$,\foot{ corresponding to the
one relevant direction, all $\mu_p$ dependence being absorbed in $g_p$}\
it merely serves to fix the physical
mass scale $m_p$. If we provide a physical definition for this mass-scale
(equivalent to a renormalization scheme for $g_p$) then
the dimensionless coupling constant ratios
 formed using powers of $m_p$, 
will be universal. Evidently these are generated by  
the dimensionless potential (and field):
\eqn\unip{U({\vphi})=V_p(m_p^d{\vphi},0)/m_p^D\quad,}
from which it follows that the equation
$$U'({\vphi})=h\quad,$$ 
is nothing but the universal
scaling equation of state ($h$ being the external `magnetic field', equivalent
to the source $J$ in quantum field theory).
Let us define a (dimensionless) running  
`mass' $m\equiv m(g)$ with the property that it
corresponds in the limit to the physical mass $m_p$, \ie
\eqn\limm{m_p=\lim_{\Lambda\to0}\Lambda\, m(g)\quad.}
It will be helpful to change variables from $g$ to $m$.
Indeed, defining
\eqn\runU{U(\vphi,m)=V(m^d\vphi,t)/m^D\quad,}
and using \sca, \limm\ and \unip, we see that 
$U(\vphi,m)$
has the property that it tends to the universal effective potential
\unip\ in the limit $m\to\infty$.
By rewriting the RG \shar\ as a flow equation for $U(\vphi,m)$,
we obtain the universal physics directly as the large $m$ behaviour.
The flow with respect to 
$t$ is never needed
explicitly, having been `hidden' in the beta-function for $m$: 
 ${\partial m/\partial t}=\beta(m)$.  

In the next three sections we flesh out the above sketch of changes of 
variables for the cases of LPA with sharp cutoff, LPA -- equivalently
$O(\partial^0)$ -- for the smooth cutoff employed in \deriv, and 
$O(\partial^2)$ for this smooth cutoff.
 In all cases we study only the symmetric phase and use as
definition of $m_p$, 
\eqn\mp{m_p^2\equiv V_p''(0,0)\quad,}
in concord with  refs.\GZJ\wet\sok. Then by \unip, $U(\vphi)$ is normalised
as \eqn\Uni{U''(0)=1\quad.}

Again, the conclusions of this section are not particular to the LPA, and
can be expected to hold also for the exact RG, with one proviso: beyond
the LPA we would have to consider the effects also of wavefunction
renormalisation. This is straightforwardly incorporated by a further
change of variables (multiplicative renormalization) to a field with
a kinetic term that remains correctly normalised under flow in $t$ (or $m$).
As a byproduct one obtains a \nonp\ expression for the running 
anomalous scaling $\gamma(m)$.
We will provide the details in sect.7, where the $O(\partial^2)$
approximation will be developed.

\newsec{Estimates from the LPA of the Wegner-Houghton RG}

\seclab\WH
In this section we carry out explicitly the program outlined above, for
equation \shar, the result of the LPA applied to the RG for sharp cutoff.
 We will also explain our numerical methods for solving the
resulting partial differential equations to obtain the universal 
coupling constant ratios \ie the Taylor expansion coefficients of 
$U(\vphi)$. Since these results refer to the case of the Wilson-Fisher fixed
point in three dimensions, we from now on set $D=3$, implying for
the LPA, $d=1/2$. 

Our first step is to define an appropriate running mass $m$. If we set
\eqn\sig{\sigma(t)=V''(0,t)\ins11{and (for later)}
\halpha(t)=V^{(4)}(0,t)/24\quad,}
then by \runU\ and \Uni,
\eqn\limsi{\sigma/m^2\to1\quad,}
as both tend to infinity. 
We cannot simply have $m^2=\sigma$ however, since $\sigma$ starts out
negative:
\eqn\sigs{\sigma(t=-\infty)=\sigma_*=-.46153372\cdots}
\hashas\trunc, this being $\sigma$'s fixed point value. Using this
in eqn.\shar, the Taylor expansion of $V(\phi)$ can be developed and
hence of course fixed point values for $\halpha$ and all the 
other couplings may
be deduced. Some of these were quoted in 
ref.\ref\bag{C. Bagnuls and C. Bervillier, Phys. Rev. B41 (1990) 402.}.
Since there has been confusion in some of the recent literature, it is
worthwhile to stress that these numbers have nothing directly to do with
the universal ratios in the scaling equation of state. In particular, fixed 
point values depend sensitively on the form of the cutoff; they are 
not universal.

Expanding \shar\ in powers of the field,
and using \sig, we have
\eqn\bets{
{\partial\sigma\over\partial t}=2\sigma+{24\halpha\over1+\sigma}\quad.}
Clearly in general, on expanding \shar, inverse powers of $1+\sigma$ are
generated, so to keep the algebra clean it is helpful to define
\eqn\shm{\sigma=-1+m^2\quad,}
in agreement with \limsi. At the same time this ensures
that $m$ is real, since from \shar\ we clearly have
$\sigma(t)>-1$ for all $t$.
Now define the couplings in $U(\vphi,m)$ by,
\eqn\als{U(\vphi,m)={1\over2} 
{\sigma\over m^2} \vphi^2+U_{int}(\vphi,m)\ins{.5}{.5}{with}
U_{int}(\vphi,m)=\E+\sum_{k=2}^\infty\vphi^{2k}\alpha_{2k}(m)\quad. }
Thus, from \sig\ and \runU\ we have $\alpha_4=\halpha/m$. 
Differentiating \runU\ with respect to $t$, and using this relation,   
\bets, \shm\ and \shar, we obtain
\eqn\shUm{{\beta}{\partial\over\partial m}U(\vphi,m)=
\left({1\over m^2}-{12\alpha_4\over m^3}\right)
\left(3U-{1\over2}\vphi U'\right)+{1\over m^3}\ln(1+U''_{int})\quad,}
with the beta function,
\eqn\shbet{\beta(m)=m-1/m+12\alpha_4(m)/m^2\quad.}
The $\vphi^2$ part of \shUm\ is satisfied automatically by \shm, 
while all other couplings, including $\alpha_4$, are
 deduced as a function of $m$ from
\shUm\ once the boundary condition at $m=m_*=\sqrt{1+\sigma_*}$ is
provided. This boundary condition may be deduced as follows. First we
need to choose a normalization for $v(\phi)$ (cf. sect. 3). We choose,
\eqn\vnorm{v''(0)=1\quad,} which by \close\ implies for small $g(t)$ 
(\ie $t\to-\infty$), $\sigma=\sigma_*+g(t)$, and thus at the fixed point
$\partial g(t)/\partial m=2m_*$.
Defining, analagously to \runU, 
\eqn\uf{U_*(\vphi)=V(\vphi\sqrt{m_*})/m_*^3\ins11{and}
u(\vphi)=v(\vphi\sqrt{m_*})/m_*^3\quad,}
substituting \close\ into \runU\ and differentiating with respect to $m$
gives the boundary condition for  $\partial U/\partial m$,
\eqn\bc{{\partial U\over\partial m}\Big|_*={1\over m_*}
(\half\vphi U'_* -3U_*) +2m_*u\quad.}
Note that $\partial U/\partial m$ needs to be specified as well as $U$
at the fixed point, 
because $\beta(m_*)=0$ and the right hand side of \shUm\ vanish there.
[Thus $m=m_*$, $U=U_*$ corresponds to a so-called `singular point' for
the differential equation \shUm.] 

Since our goal is the numerical computation 
of as many of the universal coupling constant ratios 
$\alpha_{2k}(\infty)$ [$k=2,3,\cdots$, \cf eqn.\unip\ and below \runU]
as possible, it makes sense
to Taylor expand the differential equation \shUm\ and boundary condition
\bc\ in terms of $\vphi$ and obtain directly ordinary differential
 flow equations for the $\alpha_{2k}(m)$. These are easily obtained
to high order using algebraic computing packages. (We used Maple.)
We can expect to obtain better estimates this way and with less effort 
compared to, say, numerically
solving the partial differential equations for $U(\vphi,m)$,
and then obtaining the $\alpha_{2k}(\infty)$ by
numerically differentiating $U(\vphi,\infty)$, $2k$
times at $\vphi=0$. But we now have a problem of truncation:
the expansion of \shUm\ leads to differential equations
where the $\partial\alpha_{2k}/\partial m$ depend on all $\alpha_{2j}$
up to and including $\alpha_{2k+2}$, and therefore if we keep only a 
finite number of these equations, say up to and including the
equation  $\partial\alpha_{2k}/\partial m=\cdots$,
we will require an ansatz for the
highest $\vphi$-power, the `top coupling' $\alpha_{2k+2}(m)$. 
However, this problem is not
as severe as for computations of fixed points by truncations in powers
of the field\trunc, in particular since we already know the initial (fixed
point) data numerically there will be no spurious solutions\trunc,
and we can expect that the results converge as 
we keep more and more of the equations (and thus at the same time
compute estimates for more and more of the couplings) providing
we supply a decent ansatz for the top coupling. Of course if our
ansatz corresponded exactly to the right answer for the top coupling, all
the lower couplings would be computed correctly. Our assumption is
however that if a reasonable model of the top coupling's flow is
inserted, then the lower couplings flow becomes more and more insensitive
to the error in the modelling of the top coupling as $k$ is increased
(the error having to feed down successively through more equations).
We find that the results bear out this assumption for several different
ans\"atze, at least to sufficient
accuracy for this exercise. These ans\"atze are outlined below. First 
we sketch the final steps necessary to obtain
appropriate boundary conditions for these flow equations. 
\eqnn\anzi\eqnn\bes\

From the numerical value
for $\sigma_*$, we deduce as many of the couplings in $V(\phi)$ as needed
(\cf \sigs\ and below it), and thus from \shm\ and \uf, as many of
the initial
values, $\alpha_{2k}(m_*)$, as needed. Similarly, by Taylor expansion
of \sharper, using the normalization \vnorm, and
the relevant eigenvalue $\lambda=1.450416(1)$ (already obtained
numerically in the
process of writing refs.\trunc\deriv) we can obtain the Taylor
series coefficients of $v(\phi)$. From \uf\ and \bc, this then gives the
numerical values for the initial gradients
$\alpha'_{2k}(m_*)$. Some numerical methods can cope
with this singular point [\cf below \bc], but we found that the 
resulting differential equations were remarkably well-behaved numerically
and that such sophistication was not necessary. Instead we used the 
basic fourth-fifth order Runge-Kutta provided within the Maple package,
and set the boundary condition a little way away from the fixed point
as $\alpha_{2k}(m_*+\delta)=\alpha_{2k}(m_*)+\alpha'_{2k}(m_*)\,\delta$. 
Here, $\delta$ could be set very small. We chose typically to set it to 
$\sim 10^{-6}$. The results were entirely insensitive to the precise 
choice. We integrated out to $m=20$: with the improved estimate \bes\
described below this gave us more than sufficient accuracy while being easily
achieved by the above package.

Turning now to the ansatz for the top coupling, 
the simplest workable ansatz of all, is
to ignore its evolution and just set it equal
to its initial value: 
$$\alpha_{2k+2}(m)=\alpha_{2k+2}(m_*)\quad.\eqno\anzi$$
 Numerically we found that this ansatz
results in reasonable convergence, but it is easy to do better. From the above,
we know also the initial gradient, while from the previous analysis (and 
borne out \eg by the above ansatz) we expect that all couplings 
$\alpha_{2j}(m)$
`run' with $m$ initially and then freeze out at some value $m_{freeze}$,
after which $\alpha_{2j}(m)\approx \alpha_{2j}(\infty)$. Indeed from
\shUm\ we see that asymptotically $\alpha_{2j}(m)\sim 
(1-{3-k\over 2 m^2})\alpha_{2j}(\infty)$, a fact which we used to improve the
estimate of $\alpha_{2j}(\infty)$: 
$$\alpha_{2j}(\infty)\sim \left(1+{3-k\over2m^2}\right)\alpha_{2j}(m)
+O\left({1\over m^3}\right)\quad. \eqno\bes$$
This suggests the following simple ansatz for the top coupling,
\eqn\topa{\alpha_{2k+2}(m)=\alpha_{2k+2}(m_*)+
{\alpha_4(m)-\alpha_4(m_*)\over\alpha'_4(m_*)}\alpha'_{2k+2}(m_*)\quad,}
which thus has the right value and gradient at $m=m_*$ and will freeze out
at the same value $m_{freeze}$ as $\alpha_4$. Initially we employed
an ansatz where $\alpha_4$ was replaced in \topa\ by $\alpha_{2k}$, in the
expectation that any $k$ dependence of $m_{freeze}$ would thus be better
taken into account:
\eqn\anzii{\alpha_{2k+2}(m)=\alpha_{2k+2}(m_*)+
{\alpha_{2k}(m)-\alpha_{2k}(m_*)\over\alpha'_{2k}(m_*)}\alpha'_{2k+2}(m_*)\quad.}
 It is such an ansatz that led to the $O(\partial^2)$
estimates quoted in ref.\GZJ. However, we found that there was little or
no drift of $m_{freeze}$ with $k$, while using $\alpha_{2k}$ in place
of $\alpha_4$ in \topa\ led sometimes to numerical instabilities \eg when
$\alpha'_{2k}(m_*)$ happened to be much smaller than $\alpha'_{2k+2}(m_*)$,
and to poorer convergence.

All three methods described led to consistent estimates for the 
$\alpha_{2k}(\infty)$. We present some of the raw data
in sec.8. The most convergent answers
 were produced with ansatz \topa. Using these, and untying the numerical
transformation \pret, \ie as
\eqn\unty{\alphap_{2k}=\zeta^{2k-2}\alpha_{2k}(\infty)\quad,}
where by below \pret, $\zeta=2\pi$, yields
our final results. These are
given in table 7. They correspond to the normalization
used in ref.\sok.

\newsec{Lowest order in the derivative expansion}

\seclab\Odo
In this section we carry out the program for the LPA of the RG 
written directly in terms of the Legendre effective action, \cf \Leg,
with respect to a smooth cutoff.
This corresponds to $O(\partial^0)$
in a derivative expansion of $\Gamma_\Lambda$:
\eqn\gappr{\Gamma_\Lambda[\phi]=\int\!\!d^3\!x\,
\left\{\half(\partial_\mu\phi_{\rm phys})^2
+V_{\rm phys}(\phi_{\rm phys},\Lambda)\right\}\quad,}
with no other approximation made. (We drop the ${}^c$ superscript.)
The relevant flow equations were derived
in ref.\deriv, where details of the form of the cutoff \etc can be
found. Since the critical exponents were found\deriv\ to be better 
estimated by the smooth cutoff LPA compared to the sharp case described
in the previous section, one might expect the universal couplings 
also to be better estimated. The results bear out these expectations.

Instead of the flow equation \shar, one finds\deriv:
\eqn\Vo{{\partial\over\partial t} V(\phi,t)+\half\phi V'(\phi,t)
-3V(\phi,t)=-1/\sqrt{2+V''(\phi,t)}\quad.}
We define again $\sigma(t)$ and $\halpha(t)$ by eqn.\sig. However, 
Taylor expanding \Vo\ one finds instead of \bets,
$${\partial\sigma\over\partial t}=2\sigma 
+{12\,\halpha\,(2+\sigma)^{-3/2}}\quad,$$
the difference of course arising as a result of the inherently different
`blocking' procedure implied by use of the sharp cutoff.
(Only universal quantities such as the exponents and the 
$\alpha_{2k}(\infty)$ are independent of the choice of cutoff
and this only when the RG is computed exactly, otherwise some dependence
on the choice of cutoff is left\revii. See also our comments below
\sigs.) By similar reasoning to that for \shm, a convenient definition
for $m$ is here,
\eqn\smm{\sigma=-2+m^2\quad.}
Using the same expansion \als, we thus obtain the flow equation
\eqn\smUm{{\beta}{\partial\over\partial m} U(\vphi,m)=
\left({2\over m^2}-{6\alpha_4\over m^4}\right)\left(3U-{1\over2}\vphi U\right)
-{1\over m^4\sqrt{1+U''_{int}}}\quad,}
with the beta-function,
$$\beta(m)=m-2/m+6\alpha_4(m)/m^3\quad.$$

Choosing the normalization \vnorm\ for the relevant perturbation,
and using \close,
\smm\ again implies $\partial g(t)/\partial m=2m_*$ as $t\to-\infty$, 
and thus with
definitions \uf, the same formula \bc\ holds for the 
initial boundary conditions. Our treatment of these and the choice of
ansatz for the top coupling are the same as in the previous section.
From \smUm, the analogous formula to \bes\ reads however,
\eqn\bess{\alpha_{2j}(\infty)\sim \left(1+
{3-k\over m^2}\right)\alpha_{2j}(m)+O\left(
{1\over m^4}\right)\quad.}
The numerical values for the initial couplings follow from\deriv\
$\sigma_*=-.534648257\cdots$, and the relevant eigenvalue
$\lambda=1.514260\cdots$. Eqn.\Vo\ was also prettified by a numerical
transformation of the form \pret, and thus must be untied as in \unty, but
the corresponding value for $\zeta$ is $\zeta=\sqrt{2\pi}$ \deriv.
Numerical results for $\alpha_{2k}^{\rm p}$ are given in sec.8.

\newsec{Order derivative-squared}

\seclab\Odt
We use the smooth cutoff Legendre flow equation of the previous section, but
improve the approximation by keeping also all terms of $O(\partial^2)$, \ie
we replace \gappr\ with
\eqn\gappro{\Gamma_\Lambda[\phi]=\int\!\!d^3\!x\,
\left\{\half(\partial_\mu
\phi_{\rm phys})^2 K_{\rm phys}(\phi_{\rm phys},\Lambda)
+V_{\rm phys}(\phi_{\rm phys},\Lambda)\right\}\quad.}
No further approximation is made. The resulting flow equations were
derived in ref.\deriv:  \eqna\seco
$$\eqalignno{&\phantom{\hbox{and}\hskip 1cm}
{\partial V\over\partial t}+{1\over2}(1+\eta)\phi V'-3V=
-{1-\eta/4\over\sqrt{K}\sqrt{V''+2\sqrt{K}} } &\seco a\cr
&\hbox{and}\hskip 1cm
{\partial K\over \partial t}+{1\over2}(1+\eta)\phi K' +\eta K=
\left(1-{\eta\over4}\right)\Biggl\{ 
{1\over48}{24KK''-19(K')^2\over K^{3/2}(V''+2\sqrt{K})^{3/2}} &\seco b\cr
&-{1\over48}{58V'''K'\sqrt{K}+57(K')^2+(V''')^2K\over K(V''+2\sqrt{K})^{5/2}}
+{5\over12}{(V''')^2K+2V'''K'\sqrt{K}+(K')^2\over\sqrt{K}(V''+2\sqrt{K})^{7/2}}
\Biggr\}\ \ .\cr}$$
The main novelty in the theoretical construction, compared to the previous
sections, is the appearance of a non-zero anomalous dimension $\eta$. 
This is determined at the fixed point itself, as consequence of a $\phi$ 
rescaling invariance which is here preserved by the derivative expansion
as a result of the careful choice of cutoff function. We refer the
reader to refs.\deriv\revii\ for the details. They will not be important
for the ensuing analysis. The non-zero value for $\eta$  means that the
couplings have anomalous scaling dimensions. These can be taken into 
account by approprate insertions of $m^\eta$, however, with an eye
to future more general applications, we will choose the more conventional
approach and introduce wavefunction renormalisation.

The change to dimensionless quantities is given, \cf \sca\ and ref.\deriv, as
\eqn\scal{\phi=\phi_p\mu_p^{\eta/2}/\Lambda^d,\quad\quad
V(\phi,t)=V_p(\phi_p,\Lambda)/\Lambda^3\ins{.2}{.2}{and}
K(\phi,t)=(\Lambda/\mu_p)^\eta K_p(\phi_p,\Lambda)\quad,}
where now $d=\half(1+\eta)$, and here the powers of $\mu_p$
are introduced to balance engineering dimensions. These physical
quantities are related to the true physical quantities by the numerical
transformation \pret, and $K_p=K_{\rm phys}$ \deriv, with $\zeta=
\sqrt{2\pi}$ as in sec.\Odo. As previously, we will correct for this
transformation at the end of the section. 

The physical fields do not yet have a normalised kinetic term. It is
helpful to define
\eqn\zdef{z(t)=1/\sqrt{K(0,t)}\quad,}
then from \scal\ and \gappro, the renormalised physical fields 
\eqn\renp{\phi_R =\e{\eta t/2}\phi_p/z(t)}
have a normalised kinetic term
$K_R(\phi_R,\Lambda)=K_p(\phi_p,\Lambda)/K_p(0,\Lambda)$, satisfying
 $K_R(0,\Lambda)=1$. (The normalisation
of $V_R(\phi_R,\Lambda)=V_p(\phi_p,\Lambda)$ is of course unaffected.)
Now we can define the physical mass scale [\cf \mp],
\eqn\lmp{m_p^2=V''_R(0,0)\quad,}
and thus the universal dimensionless potential and $O(p^2)$ parts,
\eqn\lunip{U(\vphi)=V_R(\vphi\sqrt{m_p},0)/m_p^3\quad,\hskip1cm
L(\vphi)=K_R(\vphi\sqrt{m_p},0)\quad.}
Note that $U(\vphi)$ is still normalised as in \Uni, 
while evidently $L$ is normalised as
\eqn\Lni{L(0)=1\quad.}
Introducing again a running mass $m(t)$ with the property that it 
satisfies the limit \limm, and combining the above transformations,
we see that the quantities
\eqn\runLU{U(\vphi,m)=V(\vphi z\sqrt{m},t)/m^3\ins11{and}
L(\vphi,m)=z^2K(\vphi z\sqrt{m},t)\quad,}
thoroughly enjoy the property that they tend to their universal counterparts 
as $m\to\infty$. Defining again $\sigma(t)$ as in \sig, we
see from \runLU\ and \Uni, that \limsi\ is replaced by the condition
\eqn\ulimsi{\sigma z^2/m^2\to 1\quad.}
Since $K(\phi,t)$ starts out normalised 
at the fixed point, \ie $K(0,-\infty)=K(0)=1$ \deriv, 
we have by \zdef, the initial condition $z(t=-\infty)=z_*=1$. On the other
hand $\sigma(t=-\infty)=\sigma_*$ and
we found numerically (in the process of  writing ref.\deriv),
 $\sigma_*=-.3781684\cdots$, so once again
replacing the limit by equality in \ulimsi\ will not do.
By a similar analysis to that leading to \frema\ and \smalL, we may confirm
that $K_p(\phi_p,\Lambda)$ has a finite limit as $\Lambda\to0$, and
thus determine from \scal\ that $z\sim m^{\eta/2}$ as $m\to\infty$.
Since $\eta$ is numerically small, $\eta=.05393208\cdots$ \deriv,
the choice
\eqn\smom{m^2=\sigma z^2 +2z}
satisfies \ulimsi\ while, being proportional to $\sigma+2/z$, neatening
up the small field expansion of eqns\seco{}. Now,
defining the couplings similarly to \als,  \eqna\algas
$$\eqalignno{U(\vphi,m)&={1\over2} 
{\sigma z^2\over m^2} \vphi^2+U_{int}(\vphi,m)\quad,
U_{int}(\vphi,m)=\E+\sum_{k=2}^\infty\vphi^{2k}\alpha_{2k}(m)\quad, &\algas a\cr
L(\vphi,m)&=1+\sum_{k=1}^\infty\vphi^{2k}\gamma_{2k}(m)\quad, &\algas b\cr}$$
and substituting \runLU\ into eqns\seco{}, we obtain
\eqna\secul
$$\eqalignno{
&\beta{\partial\over\partial m} U(\vphi,m) = {\beta-m\over m}
\left({1\over2}\vphi U'-3U\right)+
{\gamma-\eta\over2}\vphi U'
-\left(1-{\eta\over4}\right){z^2\over m^4R\sqrt{L}}\hskip.5cm 
\hskip1cm &\secul a\cr
&\beta{\partial\over \partial m} L(\vphi,m)=(\gamma-\eta)
\left({1\over2}\vphi L'+L\right)+{\beta-m\over2m}\vphi L' 
+\left(1-{\eta\over4}\right)\Biggl\{ {z^2\over48}{24LL''-19(L')^2\over 
  m^4R^3L^{3/2}}\hskip2cm  &\secul b\cr
&-{z\over48}{58m^2zU'''L'\sqrt{L}+57z^2(L')^2+m^4(U''')^2L\over m^6R^5L}
+{5z^2\over12}{m^4(U''')^2L+2m^2zU'''L'\sqrt{L}+z^2(L')^2\over m^8R^7\sqrt{L}}
\Biggr\},\cr}$$
where we have introduced the short-hand,
$$R=\sqrt{1+U''_{int}+{2z\over m^2}\left(\sqrt{L}-1\right)}\quad,$$
and defined the beta function $\beta(m)={\partial m/ \partial t}$ 
and running anomalous dimension,
$$\gamma(m)={2\over z}
{\partial z\over \partial t}\quad.$$
The factor of 2 here corresponds to the usual definition in terms of a
replacement $z=\sqrt{Z}$ in \zdef\ and \renp. These functions may be derived
analagously to that of \bets--\shbet, or directly by evaluating the 
$\vphi^2$ coefficients in \secul{a}, and the constant term in \secul{b},
using \algas{}. Thus,
\eqn\begam{\eqalign{\beta &=m+\left({\eta\over2}-2\right){z\over m}+
\left(1-{\eta\over4}\right)\left(6{z^2\alpha_4\over m^3}
+{z^3\gamma_2\over m^5}\right)\cr
\gamma&=\eta-\left(1-{\eta\over4}\right){z^2\gamma_2\over m^4}\quad.\cr}}

To determine the initial boundary conditions we again use \close, together
with the $K$ component: 
\eqn\closeK{K(\phi,t)=K(\phi)+g(t) k(\phi)\quad,} where $k(\phi)$ 
is the $O(\partial^2)$ part of the relevant operator \deriv. It is simpler
to implement the normalisation $k(0)=1$, which thus replaces \vnorm.
If we define for convenience, $\tau=v''(0)$, then \close, \closeK,
\sig\ and \zdef\ imply that for $g(t)$ small, \ie $t\to-\infty$, 
$$\sigma(t)=\sigma_*+\tau g(t)\ins11{and} 1/z^2(t)=1+g(t)\quad.$$
Substituting \smom\ and differentiating with respect to $m$ we deduce
that at the fixed point,
$${\partial g\over\partial m}=-2{\partial z\over\partial m}=
{2m_*\over1+\tau-m_*^2}\quad.$$
Thus, using the same definition \uf\ (because $z_*=1$), and analogously
$$L_*(\vphi)=K(\vphi\sqrt{m_*})\ins11, l(\vphi)=k(\vphi\sqrt{m_*})\quad,$$
combining \close, \closeK, and \runLU, and differentiating with respect to
$m$ gives the boundary conditions for the gradients,
$$\eqalign{{\partial U\over\partial m}\Big|_* &={1\over m_*}
\left({1\over2}\vphi U'-3U\right)+{m_*\over m^2_*-1-\tau}(\vphi U'-2u)\cr
{\partial L\over\partial m}\Big|_* &={1\over 2 m_*}\vphi L'
+{m_*\over m^2_*-1-\tau}(2L+\vphi L'-2l) \quad.\cr}$$

As before, we perform a small-field expansion of these equations, 
determining the initial data for the couplings in \algas{}, in the
way described in sec.\WH, using the numerical
values of $\sigma_*$ and $\eta$
already given, and (from the analysis that led to ref.\deriv)
$\lambda=1.61796660\cdots$, $\tau=-24.472671\cdots$.

Expanding \secul{}\ in powers of $\vphi$, one sees that the flow equations
for the couplings have the following structure: two infinite sequences
of equations, the differential equations for $\alpha_{2k}$,
with $k=2,3,\cdots$,
 and the $\gamma_{2k}$ equations with $k=1,2,\cdots$, each of form,
$$\eqalign{{\partial\over \partial m}\alpha_{2k}(m) &=
f^\alpha_{2k}(\alpha_4,\cdots,\alpha_{2k+2};\gamma_2,\cdots,\gamma_{2k})
\quad,\cr
{\partial\over \partial m}\gamma_{2k}(m) &= 
f^\gamma_{2k}(\alpha_4,\cdots,\alpha_{2k+2};\gamma_2,\cdots,\gamma_{2k+2})
\quad,\cr}$$
(for some functions $f^\alpha_{2k}$, $f^\gamma_{2k}$). 
This structure implies that, for given $k\ge2$,
 there is not one unique minimal way to close the equations, rather there
are two. These are:
\vfill\eject

\item{i)} The differential equations for $\alpha_4,\cdots,\alpha_{2k}$
and  $\gamma_2,\cdots,\gamma_{2k-2}$ are kept,  with ans\"atze being
supplied for $\gamma_{2k}$ and $\alpha_{2k+2}$.

\item{ii)} The differential equations for $\alpha_4,\cdots,\alpha_{2k}$
and $\gamma_2,\cdots,\gamma_{2k}$ are kept, with ans\"atze being
supplied for $\gamma_{2k+2}$ and $\alpha_{2k+2}$.

Since there is no {\it a priori} reason to prefer one method over the other,
we use both. Indeed, by employing at each $k$, first method (i) and then
method (ii), we increment sequentially by one each time, 
the number of couplings
which are estimated. For those top couplings that require an ansatz we
continue to use \topa, with also $\alpha_{2k+2}$ replaced by $\gamma_{2k}$ or
 $\gamma_{2k+2}$ as appropriate. (Of course other ans\"atze are possible.)

From a large $m$ analysis of \secul{}\ and \begam, we determine that an
improved estimate for $\alpha_{2j}(\infty)$ can be obtained as
$$\alpha_{2j}(\infty)\sim \left(1+(3-j){z(m)\over m^2}\right) 
\alpha_{2j}(m)\quad,$$
which thus replaces \bess. The $\gamma_{2j}$ have also $z/m^2$ corrections
but with a more complicated coefficient [the result of expansion of
$(U''')^2(1+U'')^{-5/2}$]. Rather than compute this coefficient algebraically,
we determined it by a simple numerical fit on neighbouring 
values of $m$ in the solution $\gamma_{2j}(m)$. This gave more than
sufficient accuracy for the improved estimate.

Finally, we have to fold back in the missing numerical factors [\cf below
\scal], thus the final results are given as
\eqn\untyg{\alphap_{2k}=\zeta^{2k-2}\alpha_{2k}(\infty)\ins11{and}
\gammap_{2k}=\zeta^{2k}\gamma_{2k}(\infty)\quad,}
where $\zeta=\sqrt{2\pi}$.

\newsec{The numerical results}

In this section we display the numerical results obtained from the analysis
described in the previous three sections. In table 1, we compare the 
results obtained for $\alphap_4$, at $O(\partial^0)$ approximation 
as in sec.\Odo, employing the three different closure ans\"atze, \anzi, \topa\ 
and \anzii\ for the `top' coupling $\alpha_{2k+2}(m)$. One sees that 
indeed there is apparent convergence for each method as $k$ is increased.
We provide also in table 1 an estimate of the converged answer for each
ansatz, together with an error which reflects the spread of results over the
several largest values of $k$ in the table.  
$$\vbox{\offinterlineskip\hrule\halign{\vrule#&&\strut\ #\ \hfil\vrule\cr
&\hfil k &\hfil\anzi\ &\hfil\anzii\ &\hfil\topa\ \cr
\noalign{\hrule}
&\hfil 2         & 1.1955 & 1.3137  & 1.3137\cr
&\hfil 3         & 1.2917 & 1.3097  & 1.3097\cr
&\hfil 4         & 1.3117 & 1.3052  & 1.3041\cr
&\hfil 5         & 1.3066 & 1.2954  & 1.3008\cr
&\hfil 6         & 1.3004 & 1.3003  & 1.3003\cr
&\hfil 7         & 1.2994 & \hfil - & 1.3010\cr
&\hfil 8         & 1.3008 & 1.3017  & 1.3014\cr
&\hfil 9         & 1.3017 & 1.3014  & 1.3014\cr
&\hfil 10        & 1.3016 & \hfil - & 1.3012\cr
\noalign{\hrule}
&\hfil $\infty$  & 1.3016(8) & 1.301(1)& 1.3012(2)\cr}\hrule}$$
Table 1. $O(\partial^0)$ results for $\alphap_4$, at different
levels $k$ of truncation, using three different closure ans\"atze.
For the two missing values in the middle
column, the integrating routine failed to converge, 
due to
wild numerical behaviour of the ansatz [\cf 
below \anzii.]

The three methods give consistent estimates, although
method \topa\ is seen to be the most powerful.  
Since we find that these facts hold true for the other couplings (and the other
flow equations), from now on we present numerical results for method 
\topa\ only.
$$\vbox{\offinterlineskip\hrule\halign{\vrule#&&\strut\ #\ \hfil\vrule\cr
&\hfil k &\hfil$\alphap_4$
&\hfil$\alphap_6$ &\hfil$\alphap_8$ &\hfil$\alphap_{10}$ 
&\hfil$\alphap_{12}$ &\hfil$\alphap_{14}$ &\hfil$\alphap_{16}$ 
&\hfil$\alphap_{18}$ &\hfil$\alphap_{20}$\cr
\noalign{\hrule}
&\hfil 2 & 1.3137 &       &       &        &       &      & & & \cr
&\hfil 3 & 1.3097 & 2.7808 &       &        &       &      & & & \cr
&\hfil 4 & 1.3041 & 2.6873 & 1.964 &        &       &      & & & \cr
&\hfil 5 & 1.3008 & 2.6556 & 1.704 & -3.566 &       &      & & & \cr
&\hfil 6 & 1.3003 & 2.6537 & 1.730 & -2.718 & -1.83 &      & & & \cr
&\hfil 7 & 1.3010 & 2.6600 & 1.779 & -2.292 &  2.05 & 17.1 & & & \cr
&\hfil 8 & 1.3014 & 2.6632 & 1.795 & -2.231 &  2.00 & 9.72 & 20.0 & & \cr
&\hfil 9 & 1.3014 & 2.6628 & 1.790 & -2.286 &  1.46 & 4.70 & -24.8 & -47.3 & \cr
&\hfil 10 &1.3012 & 2.6615 & 1.783 & -2.324 &  1.25 & 3.73 & -26.4 & -16.9 & 
                                                                    -389 \cr
\noalign{\hrule}
&\hfil $\infty$  &
         1.3012(2)& 2.662(2) & 1.78(1)& -2.3(1) & 1.3(7)& 6(4)& -20?& -20? 
& \hfil ?\cr}\hrule}$$
Table 2. $O(\partial^0)$ results for all computed couplings 
$\alphap_{2j}$, at different levels $k$ of truncation via
\topa, together with estimates for  $k=\infty$. 

$$\vbox{
\offinterlineskip\hrule\halign{\vrule#&&\strut\ #\ \hfil\vrule\cr
&\hfil k &\hfil$\alphap_4$ &\hfil$\alphap_6$ 
&\hfil$\alphap_8$ &\hfil$\alphap_{10}$ 
&\hfil$\alphap_{12}$ \cr
\noalign{\hrule}
& 2 (i)     & .8682 &	  &	   &        &     \cr
& 2 (ii)    & .8688 &	  &	   &        &     \cr
& 3 (i)     & .8644 & 1.253  &        &        &     \cr
& 3 (ii)    & .8667 & 1.269  &        &        &     \cr
& 4 (i)     & .8639 & 1.237  & .586   &        &     \cr
& 4 (ii)    & .8646 & 1.240  & .5790  &        &     \cr
& 5 (i)     & .8639 & 1.235  & .5587  & -0.950 &     \cr
& 5 (ii)    & .8633 & 1.230  & .5275  & -1.112 &     \cr
& 6 (i)     & .8635 & 1.232  & .5475  & -0.873 & -.176\cr
& 6 (ii)    & .8630 & 1.229  & .5359  & -0.899 & -.0688\cr
\noalign{\hrule}
&\hfil $\infty$  & .8635(5) & 1.230(3)& .54(1) & -.95(5)& \hfil ?\cr}\hrule}$$
Table 3. $O(\partial^2)$ estimates of $\alphap_{2j}$
at different levels of truncation $k$ via
\topa, for the two closures described at the end of sec.\Odt, together
with estimates for $k=\infty$.

$$\vbox{\offinterlineskip\hrule\halign{\vrule#&&\strut\ #\ \hfil\vrule\cr
&\hfil k &\hfil$\gammap_2$
&\hfil$\gammap_4$ &\hfil$\gammap_6$ &\hfil$\gammap_8$ &\hfil$\gammap_{10}$ 
&\hfil$\gammap_{12}$ \cr
\noalign{\hrule}
& 2 (i)     & .6427 &	  &	   &        &    & \cr
& 2 (ii)    & .6472 & -3.023 &	   &        &     &\cr
& 3 (i)     & .6442 & -3.047 &        &        &     &\cr
& 3 (ii)    & .6537 & -2.748  & 7.26 &        &    & \cr
& 4 (i)     & .6504 & -2.782  & 7.10  &        &     &\cr
& 4 (ii)    & .6524 & -2.771  & 6.59  &  7.82  &     &\cr
& 5 (i)     & .6514 & -2.776  & 6.55  & 7.3 &     &\cr
& 5 (ii)    & .6499 & -2.801 & 6.05  & -2.5 &  -69   &\cr
& 6 (i)     & .6502 & -2.797  & 6.10  & -2.20 & -71.3 &\cr
& 6 (ii)    & .6492 & -2.805  & 6.02  & -2.4 & -53.7 & -15\cr
\noalign{\hrule}
&\hfil $\infty$  & .6497(5) & -2.805(5)&6.06(4)  &-2.3(1) &-62(9)
 & \hfil ?\cr}\hrule}$$
Table 4. $O(\partial^2)$ estimates of $\gammap_{2j}$
at different levels of truncation $k$ via
\topa, for the two closures described at the end of sec.\Odt, together
with estimates for $k=\infty$.

The estimates
resulting from LPA of the Wegner-Houghton equation are given in
table 5.
(The accuracy of these numbers was improved by correcting also for the
$1/m^3$ contributions in \bes, in a similar way to that done for $\gammap_{2j}$
in the previous section.) To our knowledge these estimates for
  universal coupling constant ratios have not
been derived before.\foot{We stress again that these do not correspond
to exact RG
fixed point couplings [\cf our discussion above \bets].}\ ({\it N.B.} this
approximation gives $\nu=.6895$ and $\omega=.5952$ \deriv\hashas.)

$$\vbox{\offinterlineskip\hrule\halign{\vrule#&&\strut\ #\ \hfil\vrule\cr
&\hfil k &\hfil$\alphap_4$ &\hfil$\alphap_6$ 
&\hfil$\alphap_8$ &\hfil$\alphap_{10}$ 
&\hfil$\alphap_{12}$ &\hfil$\alphap_{14}$ &\hfil$\alphap_{16}$ 
&\hfil$\alphap_{18}$ \cr 
\noalign{\hrule}
&\hfil 2         & 1.5308 &	  &	   &        &      & & & \cr
&\hfil 3         & 1.5288 & 3.569 &        &        &      & & & \cr
&\hfil 4         & 1.5184 & 3.439 & 2.142  &        &      & & & \cr
&\hfil 5         & 1.5107 & 3.377 & 1.833  & -5.180 &      & & & \cr
&\hfil 6         & 1.5122 & 3.396 & 2.005  & -3.491 & 5.88 & & & \cr
&\hfil 7         & 1.5163 & 3.426 & 2.142  & -2.944 & 7.30 & 14.4 & & \cr
&\hfil 8         & 1.5158 & 3.421 & 2.094  & -3.317 & 4.48 & -5.50 & -58.6 & \cr
&\hfil 9         & 1.5135 & 3.404 & 2.027  & -3.556 & 3.63 & -6.49 & -36.3 & 
31\cr
\noalign{\hrule}
&\hfil $\infty$  & 1.514(2)&3.41(1)&2.08(6)& -3.3(3)& 5(2)& -6?& -40? 
& \hfil ?\cr}\hrule}$$
Table 5. Sharp cutoff (Wegner-Houghton) with LPA,
 at different levels $k$ of truncation via
\topa, together with estimates for  $k=\infty$.

We can attempt to combine the $O(\partial^0)$ and $O(\partial^2)$ 
results of tables 2 and 3,
to give numbers with an error which takes into account
that due to truncation of the derivative expansion. Although two
terms in the derivative expansion approximation are not really enough
to take this exercise too seriously, we note that for all the results,
including exponents\deriv,\foot{For the sharp cutoff case
see above,
and refs.\deriv\hashas}\ the accuracy of the numbers obtained
(compared to results from more accurate methods  where these are available
\eg in ref.\GZJ), 
obey the following pattern: the sharp cutoff results
are the worst, $O(\partial^0)$ are better, and $O(\partial^2)$ better
still. Comparing the $O(\partial^0)$ and $O(\partial^2)$ estimates of
exponents\foot{The $O(\partial^2)$
estimate for $\omega$ in table 6,
corrects a misprint in ref.\deriv.}\ $\nu$ and $\omega$ to the worlds best determinations we find
that 1/3 of the difference between the $O(\partial^0)$ and $O(\partial^2)$
results give a good estimate of the error in the $O(\partial^2)$ results.
If we adopt this algorithm for the couplings $\alphap_{2j}$ we obtain the
numbers given in the next table. We include the data from 
ref.\deriv, and the most accurate estimates for the $\alphap_{2j}$
 -- three
dimensional resummed perturbation theory -- reported by ref.\GZJ\
[translated to this normalisation through eqn.(8.2)].
\nref\zinn{J. Zinn-Justin, 
            ``Quantum Field Theory and Critical Phenomena'' \hfil\break
(1993) Clarendon Press, Oxford.}

\midinsert
{
$$\vbox{ \offinterlineskip\hrule\halign{\vrule#&&\strut\ #\ \hfil\vrule\cr
&\hfil Approx'  &\hfil $\nu$ &\hfil $\omega$ &\hfil$\alphap_4$ 
&\hfil$\alphap_6$ &\hfil$\alphap_8$ &\hfil$\alphap_{10}$ \cr
\noalign{\hrule}
&\hfil$O(\partial^0)$& .6604 &.6285 &1.3012(2)& 2.662(2) & 1.78(1)& -2.3(1) \cr
&\hfil$O(\partial^2)$& .6181 & .8972  & .8635(5) & 1.230(3)& .54(1) & -.95(5)\cr
\noalign{\hrule}
&\hfil$\partial$ exp$^{\rm n}$ 
&.618(14) &.898(90)&.86(15)&1.2(5)&.5(4)&-1.0(5)\cr \noalign{\hrule}
& \zinn\ \& \GZJ\ &  .631(2) &.80(4) &.988(2) &1.60(1)&.83(8)&-2.0(1.3)\cr}
\hrule}$$
\centerline{\vbox{\noindent
Table 6. Estimates from the derivative expansion,
 with errors computed as described
in the text, compared to combined results from the worlds best 
determinations for the exponents \zinn\deriv, 
and resummed perturbation theory for the $\alphap$'s \GZJ.
The line labelled ``$\partial$ exp$^{\rm n}$'' gives a combined
result from $O(\partial^0)$ and $O(\partial^2)$, as explained in the
text above.}}
}
\endinsert

We have not included in table 6, the $O(\partial^2)$ estimate of $\eta$
[see above \smom], or the $O(\partial^0)$ estimates of $\alphap_{12}$
and $\alphap_{14}$ given in table 2,
because we have only one term in the derivative expansion
for these and so cannot estimate the error due to
truncation of the derivative expansion. Similarly, at $O(\partial^2)$ 
we obtain for the first time,  estimates of $\gammap_{2j}$, as
displayed in table 4. Note that the truncation level where we 
choose to stop in each of these tables, 
was determined by limitations of size and/or stability
within the Maple computing package. It is certainly possible to
do better, and thus estimate more higher order couplings,
 with a more serious attack \eg by writing FORTRAN code,
nor do we wish to rule out the possibility that 
a method based on direct integration of the
partial flow equations derived in the previous sections, could produce
competitive or better results than obtained here (while having the
advantage of course of obtaining directly also the equation of state).

It can be seen from table 6,
that while perturbative methods are more powerful than
the derivative expansion for low order couplings, the derivative expansion
eventually wins out. The reason for this is that the derivative expansion
at these lowest orders, is crude in comparison to the perturbation
theory methods, however the perturbative methods suffer from being asymptotic
 -- which in particular results in rapidly worse determinations for
higher order couplings. The derivative expansion does not suffer from this,
since it is not related at all to an expansion in powers of the field.
Indeed, it may be shown that\erg\ even at the level of the LPA,
Feynman diagrams of all topologies are included.
We recall here that derivative expansion estimates can successfully
be given for the multicritical fixed points in two dimensions\twod, where
all other standard methods fail.\foot{Needless to say, it is possible,
and would be very interesting, to
apply the methods developed here to the two-dimensional cases and thus
derive scaling equations of state for each of these multicritical points.}\
In that study, we found that derivative expansion estimates actually tend
to improve for higher dimension corrections to scaling (\ie higher 
dimension operators) and for greater multicriticality. This 
trend is opposite to that in perturbation theory (or any other standard
approximation method for that matter). This suggests that the main 
source of error in the derivative expansion estimates of the $\alphap_{2j}$'s
is that of the $\alphap_4$ coupling. We follow Guida and 
Zinn-Justin \GZJ\ and factor out the $\alphap_4$ coupling,\foot{Ref.\GZJ's
coupling $g^*\equiv24\,\alphap_4$.}\ by defining 
\eqn\f{\eqalign{ f(z) &=24\,\alpha_4(\infty)
\left\{ U\left(z/\sqrt{24\,\alpha_4(\infty)}\right)-{\cal E} \right\}\cr
 &= z^2/2+z^4/4!+\sum_{l=3} z^{2l}{F_{2l-1}/ 2l}  \quad,} }
\cf \als, \unip\ and \runU. The $F$'s are then given by
\eqn\F{F_{2l-1} =2l\,\alpha_{2l}(\infty)/\left[24\,\alpha_4(\infty)\right]^{l-1}
\ =\ 2l\,\alphap_{2l}/(24\,\alphap_4)^{l-1}\quad.}
Note that they are independent of the numerical normalisation factors
\untyg. We find much improved derivative expansion estimates of 
the higher order $F$'s compared to the $\alphap$'s. 
This is illustrated in table 7. It seems therefore that the main source of
error in the derivative expansion can be absorbed in an effective 
normalisation factor $\zeta_{eff}=\zeta \sqrt{(\alphap_4)_{exact}/\alphap_4}$.
At $O(\partial^0)$, $\zeta_{eff}=1.148(1)\,\zeta$. At $O(\partial^2)$,
$\zeta_{eff}=.935(1)\,\zeta$. We have taken the
opportunity to include in table 7, the estimates from the sharp cutoff LPA
data in table 5, and comparisons with estimates from other 
methods.\foot{We recall again that the $O(\partial^2)$ estimates 
quoted  in the corresponding table in ref.\GZJ\ were derived using the less
powerful closure ansatz \anzii. They are consistent with the present
estimates but with a larger error -- reflecting the poorer convergence
with truncation level $k$. For ref.\sok, we list only their more sophisticated
Pad\'e-Borel estimates.}
 We use the same heuristic as before to combine the $O(\partial^0)$ and
$O(\partial^2)$ results into a single derivative expansion estimate with error.
In addition, at $O(\partial^0)$ we obtain the estimates 
$F_{11}=5(3)\times10^{-7}$, $F_{13}=9(6)\times10^{-8}$, and with
 sharp cutoff LPA, the
estimate $F_{11}=9.5(4.0)\times10^{-7}$ (consistent with the 
$O(\partial^0)$ result). To our knowledge there are 
no other estimates  with which these can be compared.
\nref\rreisz{T. Reisz,{ Phys. Lett.} B360 (1995) 77.}
\nref\rZLFish{
S. Zinn, S.-N. Lai, and M.E. Fisher, {Phys. Rev. E} 54 (1996) 1176.}
\nref\rbuco{P. Butera and M. Comi, preprint IFUM 545/TH (1996).}
\nref\rkimlandau{J-K Kim and D.P. Landau, { hep-lat/9608072}.}
\vfill\eject

$$\vbox{\offinterlineskip\hrule\halign{\vrule#&&\strut\ #\ \hfil\vrule\cr
&\hfil Approx'  &\hfil$\alphap_4$ 
&\hfil$F_5$ &\hfil$F_7$ &\hfil$F_9$ \cr
\noalign{\hrule}
&\hfil Sharp &1.514(2) &.0155(3) &$3.6(5)\times10^{-4}$&
$-1.7(5)\times10^{-5}$\cr
&\hfil$O(\partial^0)$&1.3012(2)& .01638(1) & $4.68(3)\times10^{-4}$& 
$-2.4(1)\times10^{-5}$ \cr
&\hfil$O(\partial^2)$& .8635(5)& .01719(4)& $4.9(1)\times10^{-4}$ 
& $-5.2(3)\times10^{-5}$\cr
\noalign{\hrule}
&\hfil $\partial$ exp$^{\rm n}$&.86(15)&.0172(3)&$4.9(1)\times10^{-4}$&
$-5(1)\times10^{-5}$\cr
\noalign{\hrule}
&$ \hfil d=3 $~\GZJ &.988(2)&.01712(6)&$4.96(49)\times10^{-4}$&$
 -6(4)\times 10^{-5} $\cr
& \hfil   $d=3$~\sok & \hfil  & .0168 - .0173 &  4.1 - 16.2$\,\times10^{-4}$
& \cr
&$ \hfil \varepsilon{-\rm exp.}$~\GZJ & $1.2$& $ 
.0176(4) $&$  4.5(3)\times 10^{-4}  $&$
 -3.2(2) \times 10^{-5} $\cr
&$ \hfil \varepsilon{-\rm exp.}$~\sok &\hfil  & .0176 & & \cr
&\hfil ERG~\wet &$  1.20  $&$ .016 $&$  4.3\times 10^{-4} $&$   $\cr
&\hfil HT~\rreisz & .99(6) & .0205(52)&$    $&$  $\cr
&\hfil HT~\rZLFish &$1.019(6)$&$ .01780(15)$&$   $&$   $\cr
&\hfil HT~\rbuco &.987(4)&.017(1)& $5.4(6)\times10^{-4}$ &
$-2(1)\times10^{-5}$ \cr
&\hfil MC~\Tsyp &$  .97(2)  $&$ .0227(26) $&$   $&$   $\cr
&\hfil MC~\rkimlandau &$  1.020(8) $&$ .027(2) $&$  .00236(40)  $&$   $\cr
}\hrule}$$
Table 7. Estimates for $\alphap_4$ and the $F_{2l-1}$:
from the derivative expansion --
with errors computed as described in the text,
the LPA of the Wegner-Houghton equation, and (summaries of)
other methods. Again the line labelled ``$\partial$ exp$^{\rm n}$'' gives
combined results from $O(\partial^0)$ and $O(\partial^2)$, in a way
described earlier in the text.

\bigbreak\bigskip\bigskip\centerline{{\bf Acknowledgements}}\nobreak
The author would like to acknowledge 
informative correspondence with Christian Wieczerkowski.
It is a pleasure to thank
the SERC/PPARC for providing financial support through an Advanced
Fellowship, and to acknowledge support from PPARC grant GR/K55738.



 
\listrefs

\end